# An Earth-grazing fireball from the Daytime ζ-Perseid shower observed over Spain on 10 June 2012

**José M. Madiedo[1,2], Francisco Espartero[2], Alberto J. Castro-Tirado[3], Sensi Pastor[4] and José A. de los Reyes[4]**

[1] Departamento de Física Atómica, Molecular y Nuclear. Facultad de Física. Universidad de Sevilla. 41012 Sevilla, Spain.

[2] Facultad de Ciencias Experimentales, Universidad de Huelva. 21071 Huelva (Spain).

[3] Instituto de Astrofísica de Andalucía, CSIC, Apt. 3004, Camino Bajo de Huetor 50, 18080 Granada, Spain.

[4] Observatorio Astronómico de La Murta. Molina de Segura, 30500 Murcia, Spain.

[6] Universidad de Castilla-La Mancha, Toledo, Spain.

**ABSTRACT**

On 10 June 2012, an Earth-grazer meteor which lasted over 17 s with an absolute magnitude of -4.0 ± 0.5 was observed over Spain. This work focuses on the analysis of this rare event which is, to our knowledge, the faintest Earth-grazing meteor reported in the scientific literature, but also the first one belonging to a meteor shower. Thus, the orbital parameters show that the parent meteoroid belonged to the Daytime ζ-Perseid meteoroid stream. According to our calculations, the meteor was produced by a meteoroid with an initial mass ranging between 115 and 1.5 kg. During its encounter with Earth, the particle travelled about 510 km in the atmosphere. Around 260 g were destroyed in the atmosphere during the luminous phase of the event as a consequence of the ablation process. The modified orbit of the remaining material, which left our planet with a fusion crust, is also calculated.

**KEYWORDS:** meteorites, meteors, meteoroids.

# 1 INTRODUCTION

When meteoroids encounter our planet they hit the atmosphere at speeds ranging, approximately, between 12 and 72 km s$^{-1}$. As a consequence of this, most of them are completely destroyed during the ablation process before reaching the ground. However, under appropriate conditions, large and tough enough meteoroids may survive this process and reach the ground as meteorites (see e.g. Jenniskens et al. 2009, 2014). But meteoroid survival is also possible if these particles of interplanetary matter impact the atmosphere with a near-horizontal trajectory. Then, if the meteoroid is big enough and the closest height is not too low, only a portion of the incoming mass might be ablated, so that the remaining material could leave our planet with a modified orbit. These Earth-grazing events are very scarce in the scientific literature. Thus, the first Earth-grazing fireball discussed in the scientific literature was recorded on 10 August 1972 (Jacchia 1974; Ceplecha 1979, 1994), and two additional bolides recorded on 13 October 1990 and 29 March 2006 were analyzed by Borovička and Ceplecha (1992) and Abe et al. (2006), respectively.

Here we analyze an Earth-grazing meteor observed over Spain on 10 June 2012 (Figure 1). This was recorded from six meteor observing stations operating in the framework of the Spanish Meteor Network (SPMN). We also propose a new method to determine the atmospheric path of Earth-grazing events which is based on a modification of the classical planes intersection method (Ceplecha 1987).

## 2 INSTRUMENTATION AND METHODOLOGY

The Earth-grazing event discussed here was imaged by means of low-lux CCD video cameras (models Watec 902H and 902H2 Ultimate) located at sites listed in Table 1. Some of these meteor observing stations work in a fully autonomous way by means of software developed by the first author (Madiedo 2014). The cameras generate interlaced video sequences according to the CCIR video format (25 frames per second), with a resolution of 720x576 pixels. Each station employs an array of 5 to 12 cameras. Their focal length ranges from 6 to 25 mm and the field of view covered by each device ranges, approximately, from 62x50 to 14x11 degrees. A detailed description of these systems and their operation can be found, for instance, in Madiedo & Trigo-Rodríguez (2008), and Madiedo et al. (2010).

For data reduction we have employed the methods described in (Trigo-Rodríguez et al. 2009). Thus, the equatorial coordinates of the meteor along its apparent path from each station were obtained with the AMALTHEA software (Madiedo et al. 2013a,b). Then, this application was employed to calculate the atmospheric trajectory of the event. The orbital parameters of the parent meteoroid before and after its encounter with Earth were obtained by following the standard procedure described in Ceplecha (1987).

## 3 RESULTS AND DISCUSSION

The beginning of the luminous phase of the Earth-grazing event analyzed here was recorded from stations number 3 and 5 in Table 1 on 10 June 2012, at 3h 34m 55.8 ± 0.1 s UTC. The meteor moved in a SW direction and ended around 17.4 s later, at 3h 35m 13.2 ± 0.1 s UTC. The final part of the atmospheric trajectory was imaged from stations number 1, 2, and 4. The recordings show that the luminosity of the meteor was practically constant during most of the atmospheric path of the event (Figure 2). It was included in our meteor database under the code 100612. This code was given after the recording date, following the format DDMMYY, where DD is the day, MM the month, and YY the last two digits of the year.

### 3.1. Preliminary atmospheric trajectory

We employed the planes intersection method (Ceplecha 1987) to determine the atmospheric path of the fireball. This technique assumes that this trajectory is a straight line. However, this approximation holds when the length of the atmospheric path is shorter than ~100 km (Borovička and Ceplecha 1992). In this case, however, the observations indicated that this event extended over several hundred kilometers since it crossed most of Spain. As a result, the planes intersection method gave rise to a physically unrealistic result in the sense that a continuous increase in velocity was obtained by the end of this trajectory. Nevertheless, we took this result as a first approximation to the true solution. In this way, we determined that the beginning of the luminous phase of the event took place over the geographical latitude φ = 39.52 ºN and longitude λ = 1.45 ºW, at a height of 99.8 ± 0.5 km above the sea level. This position is over the province of Cuenca, in the East of Spain (at around 120 km from the Mediterranean coast). The meteoroid impacted the atmosphere with a velocity of 29.3 ± 0.3 km s⁻¹ and a zenith angle of 88.1º. In this approximation the minimum height (which corresponded to the perigee point) was of $96.2 \pm 0.5$ km at the geographical coordinates defined by $\varphi = 38.21º$ N, $\lambda = 3.81$ ºW. The ending point of the meteor was located at $\varphi = 36.90º$ N, $\lambda = 6.10$ ºW, at a height of $101.0 \pm 0.5$ km next to the zenith of the city of Lebrija, in the Southwest of Spain (at around 32 km from the Atlantic coast). Figure 2 shows the projection on the ground of the atmospheric path of the event. The total length of this trajectory was of around 510 km. The apparent radiant was located at the equatorial coordinates $\alpha = 63.6 \pm 0.1º$, $\delta = 27.42 \pm 0.05º$.

**3.2. Detailed trajectory calculation and orbital data**

Two methods have been previously proposed to determine the atmospheric path of Earth-grazing fireballs and overcome the limitations of the planes intersection method (Borovicka and Ceplecha 1992). However, they cannot be universally employed. These methods where applied to the analysis of the mag. -6 Earth-grazer fireball that was observed over Czechoslovakia and Poland on 13 October 1990. The first of these methods took advantage from the fact that the bolide passed almost through the zenith of one of the recording stations, and was seen completely on the all-sky image recorded from that location. The other method assumes that the deceleration experienced by the meteoroid during its atmospheric trajectory is negligible (Borovicka, pers. comm.). Here we propose another technique which is based on a variation of the planes intersection method and is suitable for events recorded by video devices. To understand the main idea behind this method, one should take into account that, as previously mentioned, the planes intersection method is limited to relatively short (below than ~100 km) atmospheric paths. So, if we break a long meteor path into several small segments, the planes intersection method could be successfully employed sequentially on each individual segment. In this context, each segment is considered from the mathematical point of view as an individual meteor path, in such a way that the final point of a segment corresponds exactly with the beginning point of the following one. But the way this segmentation must be done is not trivial, since it must be performed on each apparent path recorded from each observing station. And, in order to be consistent, the limiting points (i.e., the initial and final heights) of each individual segment must be exactly the same for each station. In order words, each segment must correspond to exactly the same portion of the actual atmospheric path of the meteor as seen from each recording station. The observing time t can be employed as segmentation parameter, provided that the recording systems on each station are properly synchronised. In the case considered here, a total of 14 segments have been employed (Table 2).

The data that allow obtaining the heliocentric orbit of the meteoroid before its encounter with our planet are the position of the radiant and the preatmospheric velocity, which are derived for the beginning of the atmospheric path of the meteor (Ceplecha 1987). So, if we restrict our analysis to a segment corresponding to the initial part of the meteor trajectory, we can correctly derive the value of these parameters and avoid the mathematical artifacts that arise when the planes intersection method is employed with the whole atmospheric path. We have set this initial segment by taking into consideration the recordings performed during the first 1.40 seconds of the luminous phase. In this way we have determined that the apparent radiant was located at the position $\alpha = 62.9 \pm 0.1º$, $\delta = 26.84 \pm 0.05º$, with the meteoroid striking the atmosphere with an initial velocity $V_o = 29.6 \pm 0.3$ km s$^{-1}$ at a height of $99.9 \pm 0.5$ km above the sea level (Table 3), over the point defined by $\varphi = 39.53$ ºN and longitude $\lambda = 1.43$ ºW. The inclination of this trajectory was of about 1.9 º with respect to the ground. The calculated orbital data are shown in Table 4. The orbital parameters reveal that the meteor was associated to the Daytime $\zeta$-Perseid (ZPE) meteoroid stream, which peaks around June 9 (Jenniskens 2006). Thus, the Southworth and Hawkins $D_{SH}$ dissimilarity function (Southworth & Hawkins 1963) yields 0.13. This value is below the 0.15 cut-off value usually imposed to establish a positive association (Lindblad 1971a,b).

In an analogous way, the modified orbit followed by the meteoroid after its encounter with Earth can be determined from the parameters determined at the end of its atmospheric path. Consequently, we have analyzed a segment corresponding to the final part of the luminous trajectory. This calculation was restricted to a time interval of 1.16 s at the end of this path. During this time the meteoroid travelled around 34 km in the atmosphere. In this way we have obtained that the meteoroid left our planet with a final velocity $V_f = 29.5 \pm 0.3$ km s$^{-1}$. Because of its interaction with the atmosphere, the particle left Earth with a fusion crust, like meteorites. The apparent radiant was located at the equatorial coordinates $\alpha = 63.1 \pm 0.1º$, $\delta = 26.63 \pm 0.05º$. The final point of the atmospheric trajectory was located over the position $\varphi = 36.89º$ N, $\lambda = 6.09$ ºW, at 100.4 $\pm 0.5$ km over the sea level, and the zenith angle at this stage was 92.8º. A straightforward calculation shows that both apparent radiants (beginning and end) lie in the same vertical plane, since their azimuths coincide when computed from the same geographical point. Thus, for instance, these azimuths yield around 56.05º from the initial point of the atmospheric path, and about 53.20º when computed from the final point of the luminous trajectory. From the conditions at the terminal point of the luminous part of the atmospheric trajectory we have calculated the modified orbit of the meteoroid. This has been performed by using again the standard method described in Ceplecha (1987). The resulting orbital parameters are listed in Table 4.

The same procedure was employed to analyze the portion of the trajectory between the initial and segments defined above. Thus, this portion of the apparent path was divided into 12 successive sections lasting between 1 and 1.4 seconds (Table 2). For each segment, the distance L travelled by the event was always below 42 km. The position of the apparent radiant as measured from each segment is also listed in Table 2. The azimuth of each radiant, as computed for instance from the initial point of the atmospheric path, yield again around 53.20º. This shows that, within the accuracy of the observed data, all the segments considered to analyze the trajectory lie in the same vertical plane. The averaged velocities obtained for each segment are plotted in Figure 4. According to this analysis, the minimum height (the perigee point) was located at $98.2 \pm 0.5$ km above the sea level over the geographical coordinates $\varphi = 38.31º$ N, $\lambda = 3.63$ ºW. With the scale corresponding to Figure 2, the projection on the ground of this trajectory is undistinguishable from the approximated path obtained in the previous section. The main trajectory and radiant data inferred for this event are listed in Table 3.

### 3.3 Photometric analysis: tensile strength and ablated mass

The light curve of the SPMN100612 bolide is shown in Figure 5. As can be noticed, this curve is rather smooth and the magnitude is practically constant along the whole atmospheric trajectory of the event.

For fireballs exhibiting flares along their atmospheric trajectory, the tensile strength of the parent meteoroid can be estimated by supposing that such flares take place as a consequence of the break-up of the particle due to the aerodynamic pressure (Trigo-Rodríguez & Llorca 2006, 2007). However, as the light curve plotted in Figure 5 shows, no flares were exhibited by this meteor. Nevertheless, in this case we have followed this approach to evaluate the maximum aerodynamic pressure suffered by the SPMN100612 meteoroid. This would give a lower limit for the strength of this particle. Thus, the aerodynamic strength S of a meteoroid moving at velocity v at a height were the atmospheric density is $\rho_a$ is given by means of the following equation (Bronshten 1981):

$$S = \rho_a \cdot v^2 \tag{1}$$

We have calculated the atmospheric density by following the NASA NRLMSISE-00 atmosphere model (Picone et al. 2002). In this way, the maximum aerodynamic pressure yields $(5.4 \pm 0.4) \cdot 10^3$ dyn cm$^{-2}$. As a comparison, since no flares are observed in the lightcurve (which means that the meteoroid did not experience any sudden disruption during its atmospheric path), we can conclude that this particle was tougher that the material forming weak cometary aggregates typical of large October Draconid meteoroids, which exhibit a tensile strength of about $1.4 \cdot 10^3$ dyn cm$^{-2}$, or even below (Sekanina 1985; Madiedo et al. 2013c).

The light curve can be also employed to infer the total mass ablated ($m_a$) between the beginning and the end of the luminous phase of this event. This mass is given by the equation

$$m_a = 2 \int_{t_e}^{t_b} I/(\tau v^2)\,dt \tag{2}$$

with $t_b$ and $t_e$ being, respectively, the times corresponding to the beginning and the end of the fireball. I is the measured luminosity of the bolide, which is related to the absolute magnitude M by means of the relationship

$$I = 10^{-0.4 \cdot M} \tag{3}$$

For this calculation we have employed the velocity dependent luminous efficiency ($\tau$) given by Ceplecha & McCrosky (1976). In this way the total ablated mass yields $m_a = 260 \pm 30$ g.

To calculate the mass of the meteoroid we have employed the following equation (Ceplecha 1975):

$$m = \left( \frac{2I}{\tau \sigma \rho_a K v^5} \right)^{3/2} \tag{4}$$

In Equation (4) v is the velocity of the bolide. The parameters $\tau$, $\sigma$, and $\rho_a$ are the luminous efficiency, the ablation coefficient and the atmospheric density, respectively. Again, this density was obtained from the NRLMSISE-00 atmosphere model (Picone et al. 2002) and the luminous efficiency was calculated by means of the relationships defined by Ceplecha & McCrosky (1976). The shape-density parameter K, which depends on meteoroid density $\rho_m$, shape factor A and drag coefficient $\Gamma$, is given by

$$K = \Gamma A \rho_m^{-2/3} \tag{5}$$

The values of $\sigma$ and $\rho_m$ are different for the four types of fireballs defined by Ceplecha (1988). By introducing these values into Equations (4) and (5) we have obtained an initial mass of 115 kg for type I fireballs, 16 kg for type II events, 1.5 kg for IIIA and 0.1 kg for IIIB. We have assumed spherical shape (A = 1.21) and $\Gamma$ = 0.58 (Ceplecha 1987). By taking into account the previously-calculated value for the total ablated mass, this result discards that the fireball was a type IIIB event. So, we conclude that the initial mass of the meteoroid ranged between 115 and 1.5 kg. Despite the initial mass is large, it is not surprising that the fireball was not so bright since the meteoroid moved along very low density atmospheric regions. The initial diameter of the particle ranges from 39 cm for a type I event (density 3.7 g cm$^{-3}$) to 15 cm for a type IIIA event (density 0.75 g cm$^{-3}$), being of about 25 cm for a type II event (density 2 g cm$^{-3}$). The initial mass and particle diameter can be further constrained by taking into account that, according to the analysis of the orbital elements, the meteoroid belonged to the Daytime $\zeta$-Perseid stream, which is a part of the Taurid complex. It has been postulated that the Taurid stream complex was produced as a consequence of the break up of a giant comet (with a diameter of the order of 40 km) about 20,000 years ago. The cascade disintegration of this comet gave rise to Comet 2P/Encke, which was associated with the Taurids by Whipple (Whipple 1940), but also to a large number of near-Earth objects (Clube & Napier 1982; Steel et al. 1991; Asher 1991; Asher et al. 1993; Steel & Asher 1996). In fact, Comet 2P/Encke has been proposed as the parent body of the Daytime ζ-Perseids (Jenniskens 2006). So, by assuming that the meteoroid that produced the fireball discussed here was associated with this comet and by taking into account that fireballs produced by large meteoroids from 2P/Encke have been associated to type II events (Konovalova 2003), the initial mass of this meteoroid would be 16 kg.

## 4 CONCLUSIONS

A fireball with an absolute magnitude of -4.0 ± 0.5 was observed over Spain on 10 June 2012. We have presented a new method to determine the atmospheric path of Earth-grazing events. This method, which can be employed with images taken by video devices provided that the recording systems are synchronized, overcomes the difficulties inherent to the planes intersection method, which cannot be employed to analyze these events.

According to our calculations, the meteoroid hit the atmosphere almost tangentially, with a zenith angle of about 88.1°. The total distance travelled by this particle in the atmosphere was of about 510 km. The radiant and orbital data reveal that the meteoroid belonged to the Daytime ζ-Perseid stream. From the initial meteoroid mass, which ranged between 115 and 1.5 kg, only about 260 g were ablated in the atmosphere. The most likely scenario, by assuming that the parent body of the Daytime ζ-Perseid stream is Comet 2P/Encke, is that the initial meteoroid mass is of about 16 kg. The remaining material left our planet with a fusion crust and with a modified heliocentric orbit.

## ACKNOWLEDGEMENTS

We thank *AstroHita Foundation* for its continuous support in the operation of the meteor observing station located at La Hita Astronomical Observatory. We also thank Dr. Jiri Borovička for his valuable help to improve the paper.

**TABLES**

Table 1. Geographical coordinates of the meteor observing stations involved in this work.

| Station # | Station name | Longitude (W) | Latitude (N) | Altitude (m) |
|-----------|--------------|---------------|--------------|--------------|
| 1 | Sevilla | 05º 58´ 50" | 37º 20´ 46" | 28 |
| 2 | El Arenosillo | 06º 43´ 58" | 37º 06´ 16" | 40 |
| 3 | La Hita | 03º 11' 00" | 39º 34' 06" | 674 |
| 4 | Huelva | 06º 56' 11" | 37º 15' 10" | 25 |
| 5 | Molina de Segura | 01º 09' 50" | 38º 05' 54" | 94 |
| 6 | Observatorio de Sierra Nevada | 03º 23' 05" | 37º 03' 51" | 2896 |

Table 2. Segments employed to analyze the atmospheric path of the meteor. $t_0$, $t_f$: initial and final times of each segment. L: distance travelled by the event during each segment. $H_0$, $H_f$: initial and final heights. $\alpha$, $\delta$; equatorial coordinates of the apparent radiant as measured from each segment.

| Segment # | $t_0$ (s) | $t_f$ (s) | L ±0.5 (km) | $H_0$ ±0.5 (km) | $H_f$ ±0.5 (km) | $\alpha$ ±0.1 (°) | $\delta$ ±0.05 (°) |
|-----------|-----------|-----------|-------------|-----------------|-----------------|-------------------|--------------------|
| 1 | 0 | 1.4 | 41.4 | 99.9 | 99.4 | 62.9 | 26.84 |
| 2 | 1.4 | 2.6 | 35.5 | 99.4 | 99.0 | 62.9 | 26.83 |
| 3 | 2.6 | 3.7 | 32.6 | 99.0 | 98.7 | 63.0 | 26.84 |
| 4 | 3.7 | 4.9 | 35.5 | 98.7 | 98.5 | 62.9 | 26.85 |
| 5 | 4.9 | 6.3 | 41.5 | 98.5 | 98.3 | 63.0 | 26.80 |
| 6 | 6.3 | 7.7 | 41.4 | 98.3 | 98.2 | 63.0 | 26.78 |
| 7 | 7.7 | 8.8 | 32.5 | 98.2 | 98.2 | 63.0 | 26.76 |
| 8 | 8.8 | 10.0 | 35.5 | 98.2 | 98.4 | 63.0 | 26.74 |
| 9 | 10.0 | 11.2 | 35.5 | 98.4 | 98.7 | 63.0 | 26.69 |
| 10 | 11.2 | 12.4 | 35.5 | 98.7 | 99.0 | 63.0 | 26.65 |
| 11 | 12.4 | 13.5 | 32.4 | 99.0 | 99.3 | 63.1 | 26.66 |
| 12 | 13.5 | 14.8 | 38.4 | 99.3 | 99.8 | 63.1 | 26.62 |
| 13 | 14.8 | 16.2 | 41.3 | 99.8 | 99.9 | 63.1 | 26.63 |
| 14 | 16.2 | 17.4 | 31.4 | 99.9 | 100.4 | 63.1 | 26.63 |

Table 3. Trajectory and radiant data (J2000). $H_b$, $H_m$ and $H_e$: beginning, minimum (perigee position), and ending heights of the luminous trajectory, respectively. $\alpha_{og}$ and $\delta_{og}$: right ascension and declination of the initial geocentric radiant. $\alpha_{fg}$ and $\delta_{fg}$: right ascension and declination of the final geocentric radiant. $V_o$, $V_f$: observed initial (preatmospheric) and final velocities. $V_{og}$, $V_{fg}$: geocentric velocity before and after the encounter with Earth, respectively.

| $H_b$ (km) | $H_m$ (km) | $H_e$ (km) | $\alpha_{og}$ (º) | $\delta_{og}$ (º) | $\alpha_{fg}$ (º) | $\delta_{fg}$ (º) | $V_o$ (km s$^{-1}$) | $V_{og}$ (km s$^{-1}$) | $V_f$ (km s$^{-1}$) | $V_{fg}$ (km s$^{-1}$) |
|---|---|---|---|---|---|---|---|---|---|---|
| 99.9 | 98.2 | 100.4 | 66.4 | 24.0 | 59.3 | 30.1 | 29.6 | 27.1 | 29.5 | 27.0 |
| ±0.5 | ±0.5 | ±0.5 | ±0.1 | ±0.1 | ±0.1 | ±0.1 | ±0.3 | ±0.3 | ±0.3 | ±0.3 |

Table 4. Orbital elements (J2000) of the progenitor meteoroid before and after its encounter with our planet. The nominal orbit of the ZPE stream (Sekanina 1973) has been also included.

| | a (AU) | e | i (º) | Ω (º) | ω (º) | q (AU) | P (yr) | $T_J$ |
|---|---|---|---|---|---|---|---|---|
| Before | 2.01±0.05 | 0.812±0.008 | 2.6±0.2 | 79.46538±10$^{-5}$ | 66.3±0.2 | 0.377±0.004 | 2.85±0.09 | 3.31±0.06 |
| After | 1.53±0.03 | 0.786±0.005 | 11.4±0.1 | 79.49223±10$^{-5}$ | 57.5±0.1 | 0.328±0.003 | 1.90±0.06 | 4.04±0.04 |
| ZPE | 1.918 | 0.833 | 5.3 | 78.3 | 59.2 | 0.319 | 2.65 | 3.38 |

**FIGURES**

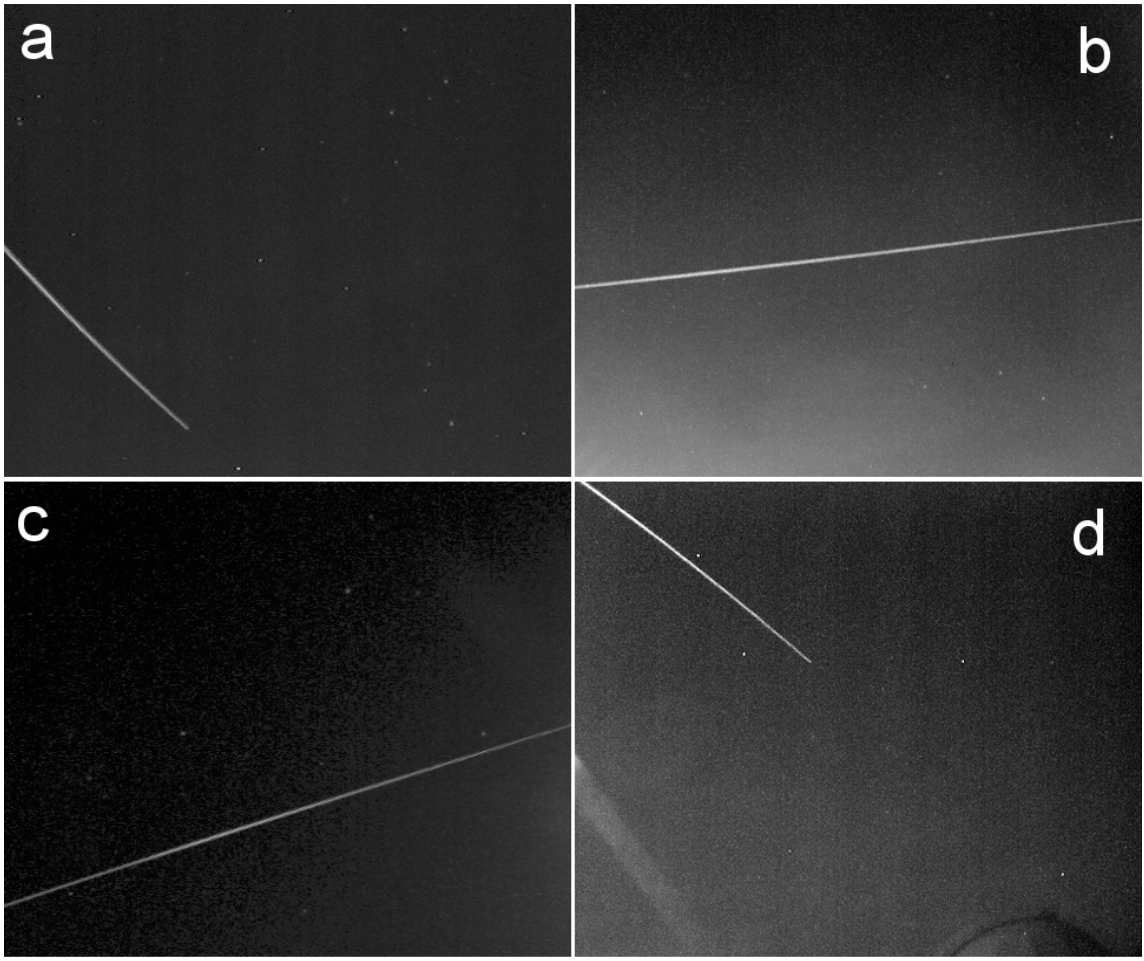

Figure 1. Sum pixel images of the 10 June 2012 Earth grazing event recorded from
Sierra Nevada (a), La Murta (b), El Arenosillo (c), and La Hita (d).

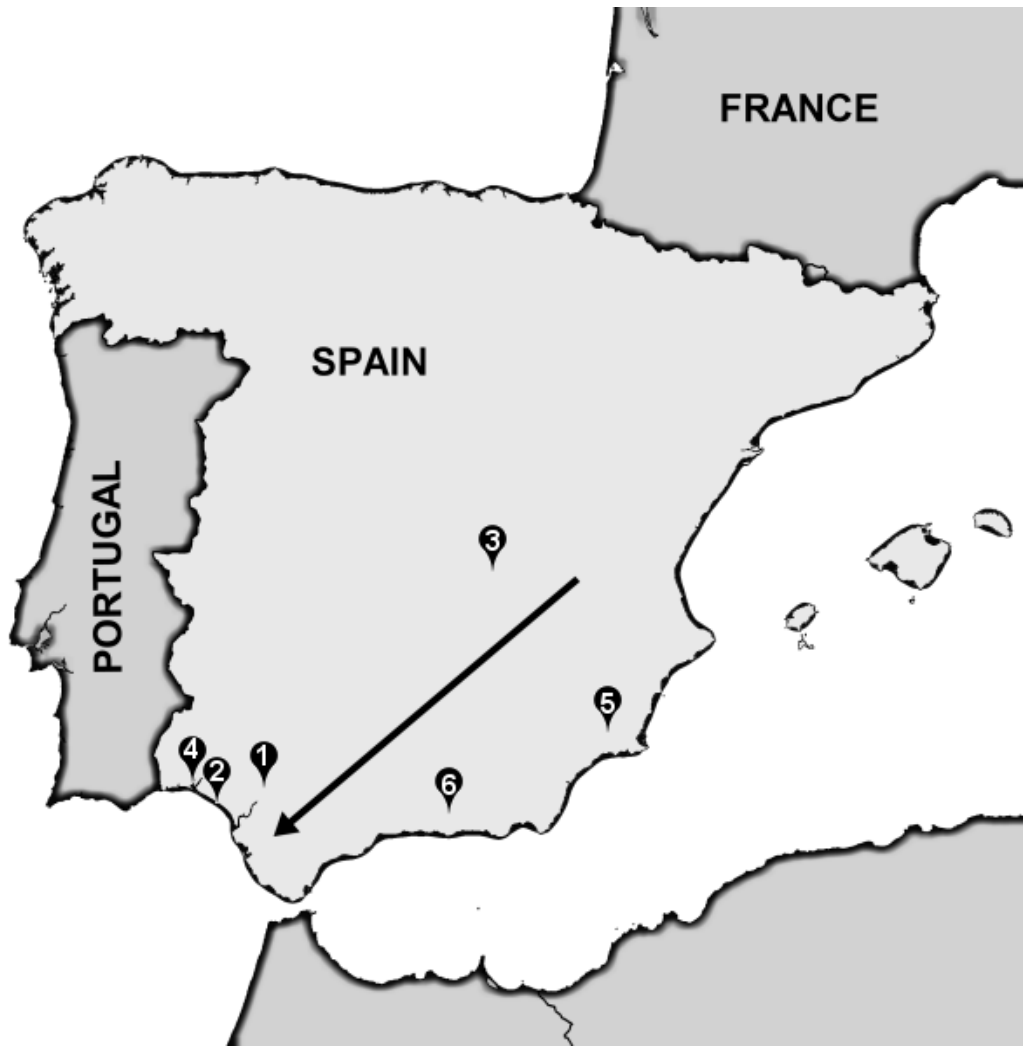

Figure 2. Projection on the ground of the luminous trajectory of the event. The position of the six meteor-observing stations listed in Table 1 has been also included.

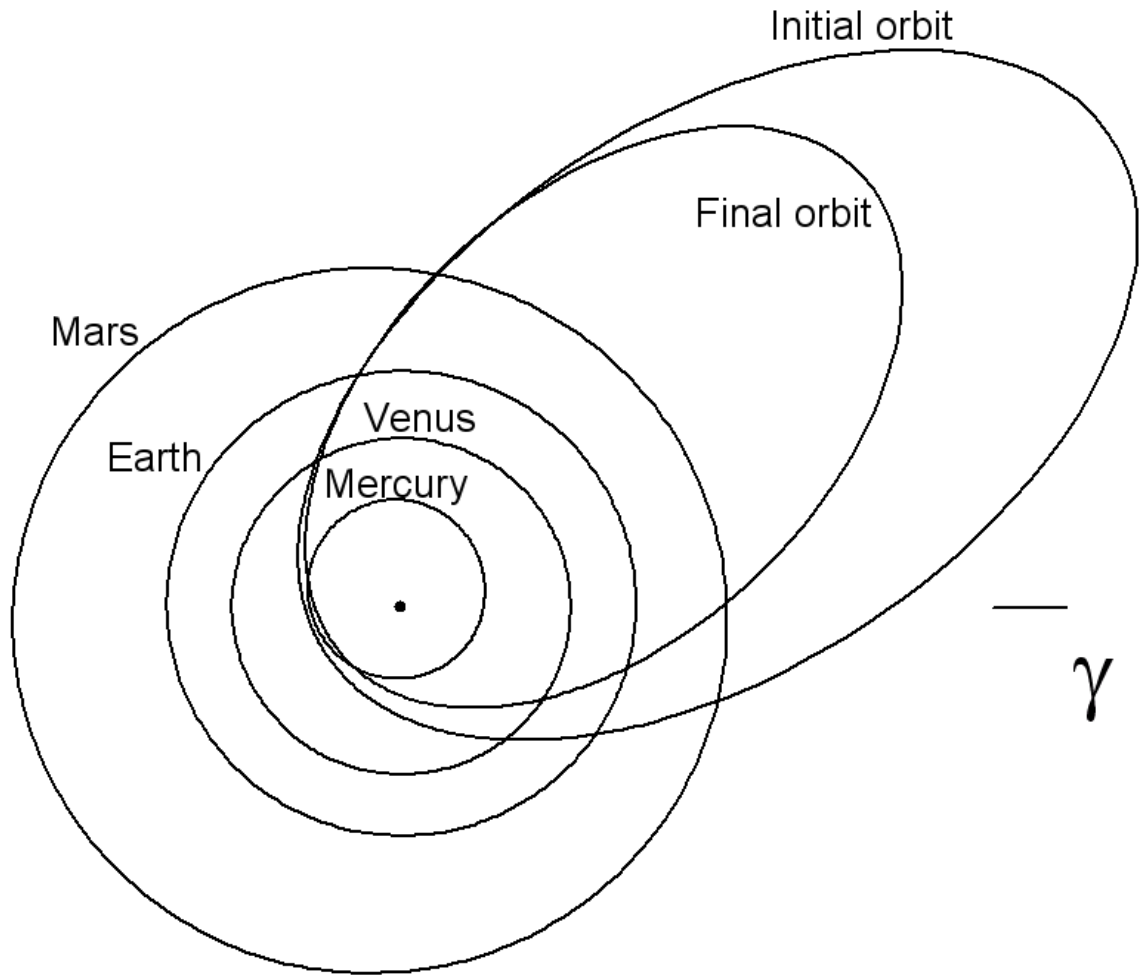

Figure 3. Projection on the ecliptic plane of the orbit of the meteoroid before and after its encounter with Earth.

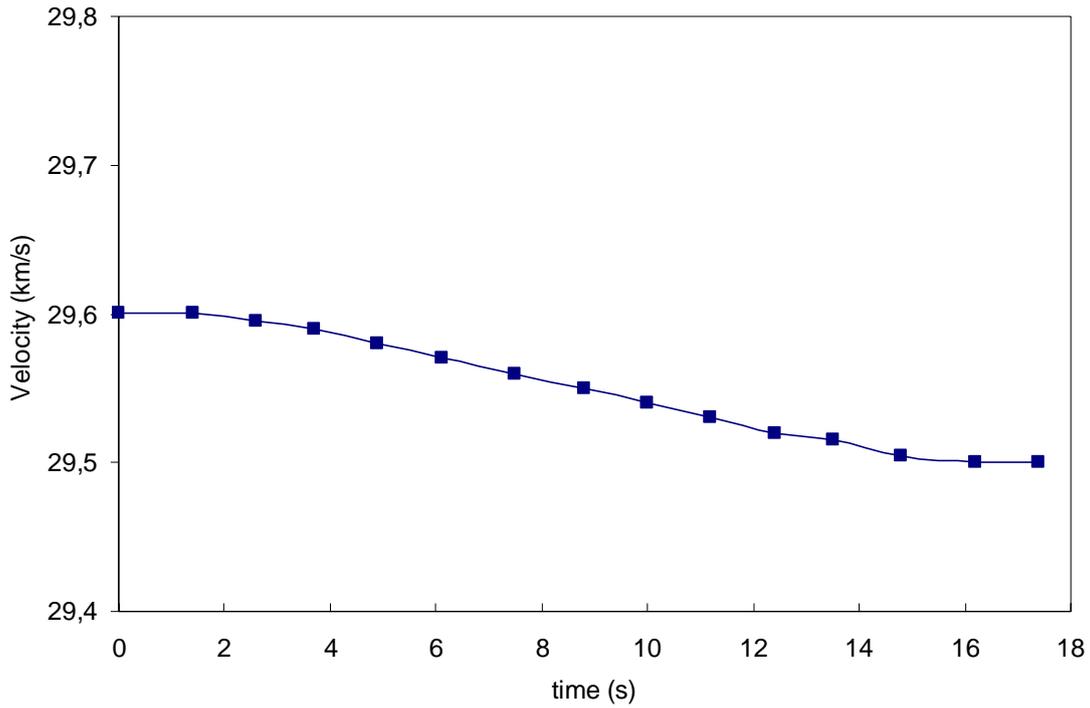

Figure 4. Meteor velocity as a function of time. Each point on this curve corresponds to the average velocity found for each of the segments in which the atmospheric trajectory was divided.

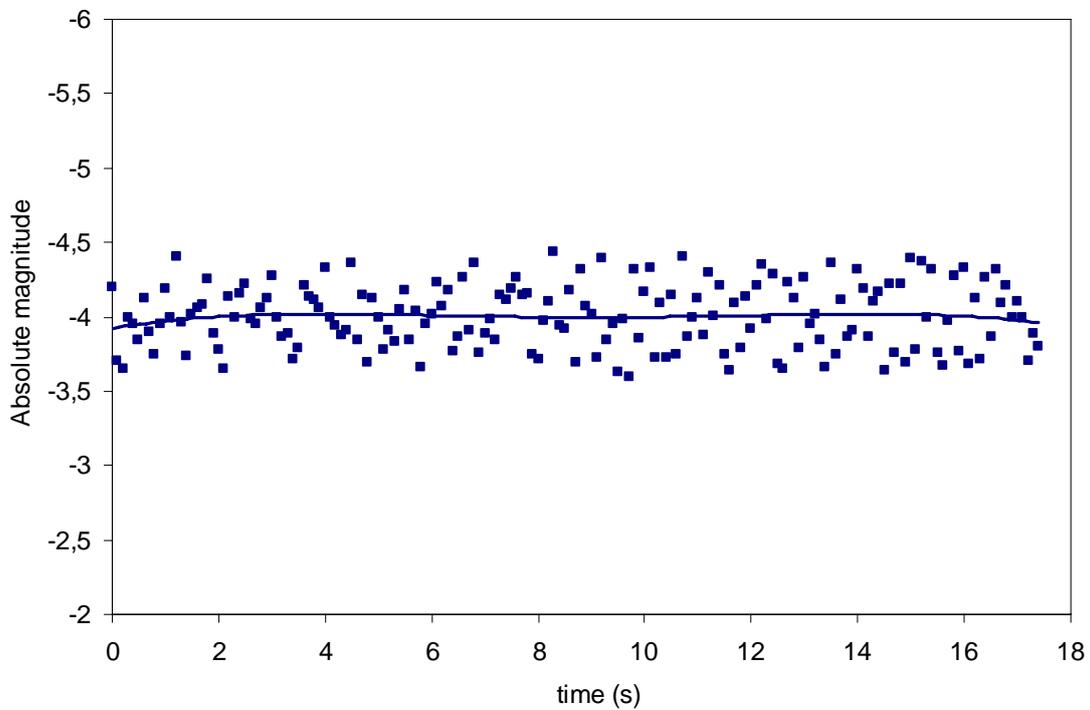

Figure 5. Lightcurve (absolute magnitude vs. time) of the meteor discussed in the text. The solid line represents smoothed data.